\input harvmac 
\input epsf

\def\Z{{\bf Z}}
\def\ie{{\it i.e.}}

\noblackbox
\Title{\vbox{\baselineskip12pt
\hbox{HUTP-99/A057}
\hbox{UKHEP/99--15}
\hbox{\tt hep-th/9910182}}}%
{\vbox{\centerline{BPS Structure of Argyres--Douglas Superconformal
Theories}}}
\bigskip
\centerline{Alfred D. Shapere\foot{shapere@pa.uky.edu}}
\smallskip
\centerline{Department of Physics and Astronomy}
\centerline{University of Kentucky}
\centerline{Lexington, KY \ \ 40502}
\bigskip
\centerline{and}
\bigskip
\centerline{Cumrun Vafa\foot{vafa@string.harvard.edu}}
\smallskip
\centerline{Jefferson Laboratory of Physics}
\centerline{Harvard University}
\centerline{Cambridge, MA\ \ 02138}

\vskip .3in

We study geometric engineering of Argyres--Douglas superconformal
theories realized by type IIB strings propagating in singular
Calabi--Yau threefolds. We use this construction to count the
degeneracy of light BPS states under small perturbations away from the
conformal point, by computing the degeneracy of D3--branes wrapped
around supersymmetric 3--cycles in the Calabi--Yau. We find finitely
many BPS states, the number of which depends on how this deformation
is done, similarly to the degeneracy of kink solutions for the
deformation of $N=2$ Landau--Ginzburg superconformal theories in two
dimensions.  Also, some aspects of worldsheet theories near general
Calabi--Yau singularities are discussed.

\Date{October 1999}

\newsec{Introduction}
In the past few years we have learned how to generate
superconformal theories in various dimensions.  Nevertheless
we do not understand the properties of conformal theories in dimensions
greater than two with the same precision as we do in the
case of two dimensions. 

In two dimensions, one powerful method to study conformal theories was
pioneered by Zamolodchikov \ref\zam{ A.B.~Zamolodchikov, ``Higher
Order Integrals Of Motion In Two-Dimensional Models Of The Field
Theory With A Broken Conformal Symmetry,'' JETP Lett. {\bf 46} (1987)
160.}, who studied their properties under special integrable massive
deformations.  Zamolodchikov's idea was to use the properties of the
deformed theories, such as the spectrum of kinks and their scattering
matrix amplitudes, to reconstruct the full data of the conformal
theory.  As a particular example of this philosophy, it was
subsequently shown in the case of massive deformations of $(2,2)$
superconformal theories in two dimensions \ref\cecva{S. Cecotti and
C. Vafa, ``On Classification of N=2 Supersymmetric Theories,'' Comm.\
Math.\ Phys. {\bf 158} (1993) 569.}\ that just the degeneracy of the
(BPS saturated) kinks of the massive theory characterizes the
dimensions of chiral operators at the conformal point.

It is thus natural, in the context of conformal theories in dimensions
bigger than 2, to also ask the same question, namely, 
what is the number of BPS states
for slight deformations of these theories away from the conformal point.
For certain cases, this has been answered.  For example
we know the degeneracy for $N=4 $ theories
in 4 dimensions \ref\sen{A. Sen, 
``Dyon--monopole Bound States, Selfdual Harmonic Forms on the 
Multi--monopole Moduli Space, and SL(2,Z) Invariance in String Theory,''
Phys.\ Lett.\ {\bf B329}, 217 (1994)
hep-th/9402032.}\ and some 
superconformal $N=2$ theories (for example $SU(2)$ with four
doublets) \ref\seiwi{N. Seiberg and E. Witten,
Nucl.\ Phys.\ {\bf B426} (1994) 19, hep-th/9407087;
Nucl.\ Phys.\ {\bf B431} (1994) 484, hep-th/9408099.}.  
The aim of this paper is to widen the class of theories for
which we know the BPS spectrum.

The case of $N=2$ field theories in 4 dimensions is already very interesting
from this point of view. In a limited set of cases, including
certain $N=2$ SU(2) gauge theories, 
the BPS spectrum can be computed exactly using field theory techniques 
\seiwi \ref\bilal{A.~Bilal and F.~Ferrari,
``The BPS spectra and superconformal points in massive N = 2
supersymmetric QCD,'' Nucl.\ Phys.\ {\bf B516} (1998) 175,
hep-th/9706145; A.~Bilal and F.~Ferrari, ``Curves of Marginal
Stability and Weak and Strong-Coupling BPS Spectra in $N=2$
Supersymmetric QCD,'' Nucl.\ Phys.\ {\bf B480} (1996)  589,
hep-th/9605101; F.~Ferrari,
``The Dyon Spectra of Finite Gauge Theories,''
Nucl.\ Phys.\ {\bf B501} (1997) 53, hep-th/9702166.}. However, 
there are many interesting 4D $N=2$ theories for
which we know very little about the structure of their BPS spectra, 
including the superconformal theories of Argyres and Douglas 
\ref\ad{P.C. Argyres and M.R. Douglas, 
``New Phenomena in SU(3) Supersymmetric Gauge Theory,'' 
Nucl.\ Phys.\ {\bf B448} (1995) 93, hep-th/9505062.}. 
In this paper, our main result will be a determination of 
the degeneracy of the BPS states under slight deformations
away from Argyres--Douglas points.\foot{A previous study of the spectrum
of an Argyres--Douglas point using $M$--theory has been made by
Gustavsson and Henningson \ref\gh{A.~Gustavsson and M.~Henningson,
``The light spectrum near the Argyres-Douglas point,''
hep-th/9906053.}.}  We realize Argyres--Douglas
superconformal theories and their deformations in terms of type IIB
strings propagating in a (nearly) singular Calabi--Yau threefold.  The
BPS states are realized in this setup in terms of D3--branes wrapped
around supersymmetric 3--cycles on the threefold.  The structure of
BPS states of Argyres--Douglas superconformal theories are among the
simplest to study from this point of view because the counting of the
supersymmetric 3-cycles gets related to counting certain special
1-cycles on a Riemann surface.  We find a structure very reminiscent
of BPS degeneracies of the deformations of minimal $N=2$ superconformal
theories in 2D. In particular we find that the spectrum of light BPS states
is finite and depends on how we deform away from the conformal point.

The structure of this paper is as follows: In section 2 we review
certain aspects of Calabi--Yau $d$-folds which have isolated
singularities.  We point out a connection between normalizable
deformations of a Calabi--Yau singularity and unitarity bounds of the
corresponding quantum field theory in the remaining ${\bf R}^n$.  In section
3 we make some remarks about the string worldsheet description near a
Calabi--Yau singularity.  In section 4 we specialize to the case of
Calabi--Yau threefold, which gives rise to $N=2$ superconformal
theories in 4 dimensions.  We set up the computation for counting the
number of supersymmetric 3-cycles for this case.  In section 5 we
apply these techniques to the case of Argyres--Douglas conformal
theories and compute their spectra.

\newsec{Calabi--Yau Singularities}

In this section we consider Calabi--Yau manifolds
with an isolated singularity.  More precisely
we consider a local model for a $d$-fold given by
a hypersurface $W(x_i)=0$ in ${\bf C}^{d+1}$.  Moreover
we assume $W(x_i)$ is a quasihomogeneous function of
$x_i$:
\eqn\quas{W(\lambda ^{q_i} x_i)=\lambda W(x_i) .}
We assume the singularity is isolated, namely
$$dW=0 \qquad {\it iff} \qquad x_i=0.$$
If $W$ is viewed as a LG superpotential of an $N=2$ theory
in $2$ dimensions 
\ref\vw{C. Vafa and N.P. Warner,``Catastrophes and the Classification of
Conformal Theories,'' Phys. Lett. {\bf B218} (1989) 51.}%
\ref\mart{E. Martinec, ``Algebraic Geometry and Effective
Lagrangians,'' Phys. Lett. {\bf B217} (1989) 431.}\
it would flow to a superconformal 
theory with central charge ${\hat c}$ given by
\eqn\centc{\hat c=\sum_{i=1}^{d+1}(1-2q_i)=(d+1)-2(\sum_i q_i). }
The condition that the Calabi--Yau threefold with 
 this singularity appear at finite distance
in moduli space was recently studied in \ref\gvw{
S. Gukov, C. Vafa, and E. Witten, ``CFT's 
from Calabi--Yau Four-folds,'' hep-th/9906070.}\
with the conclusion that if
\eqn\fd{\hat c<d-1}
the distance is finite and thus the study of singularity would be
of physical interest only in this case.

An interesting question involves deformations of the singularity.
In this context, as in the 2 dimensional case \vw\mart, it is natural
to consider the singularity ring of $W$ 
$${\cal R}={\bf C}[x_i]/dW$$
generated by the monomials 
$$x^\alpha =x_1^{\alpha_1}x_2^{\alpha_2}...x_{d+1}^{\alpha_{d+1}}$$
modulo setting to zero those polynomials which are in the ideal 
generated by $\partial_i W$.  One then considers deformations of
the form 
$$W\rightarrow W+\sum_{\alpha \in {\cal R}}g_\alpha x^{\alpha}.$$
For each element $x^\alpha$ in the ring, consider the
charge 
$$Q_\alpha=\sum_i q_i \alpha_i.$$
The $Q_\alpha$ lie in the range
 $$0\leq Q_\alpha\leq {\hat c}.$$
Moreover, the elements in the ring are paired so that
for every element of the ring with
charge $Q_\alpha$ there is one with charge $\hat c -Q_\alpha$.
The degeneracy of the ring elements is captured by
the function
$$\sum_{\alpha\in {\cal R}}t^{Q_\alpha}=
\prod_i{(1-t^{1-q_i})\over (1-t^{q_i})}$$
and the dimension of the ring ${\cal R}$ is given
by
$$N={\rm dim}\,{\cal R}=\prod_{i=1}^{d+1} {(1-q_i)\over q_i}.$$
This dimension is also the dimension of compact part of
mid-dimension homology $H_d(W=\mu)$. The compact homology
can be realized by $N$ spheres of dimension $d$ with
an intersection structure that can be obtained
from the structure of $W$ \ref\arn{V.I. Arnold, S.M. Gusein--Zade, 
A.N. Var\v encko, {\it Singularities of Differentiable Maps}
(Boston: Birkh\"auser, 1988).}.  

In questions in string theory, it is important to know how
many cohomology elements of dimension $d$ are supported on
the singularity.  
These would correspond to normalizable
$d$-forms in the internal
dimensions localized near the singularity.  
These forms can be obtained from the holomorphic $d$-form
represented as
$$\Omega ={\prod_i dx_i\over dW}.$$
What this means is that we solve the equation
$W=0$ for one of the 
$x_i$ in terms of the
others and 
view $dx_i/dW$ as $1/\partial_i W$ in the above
expression.  The other $d$-forms correspond to deformations
of $\Omega$:
$$\Omega_\alpha =\partial\Omega/\partial g_\alpha .$$
The condition that $\Omega_\alpha$ lead to a normalizable
form localized at the singularity was studied in
\gvw .  The condition for this is that $\int |\Omega_\alpha|^2$
diverge in the vicinity of $x_i=0$.  It was found that
the deformation by monomial $x_\alpha$ corresponds to
a normalizable cohomology element if
\eqn\normc{Q_\alpha < {\hat c -d+3\over 2}.}
Note that combined with \fd\ this 
implies that a normalizable cohomology deformation
must have $Q_\alpha < 1$.

The condition \normc\ can also be understood physically
by unitarity bounds of physical modes as follows.  The
coefficients $g_\alpha$ in the deformation of the singularity
should correspond in the field theory setup, either to expectation
values of scalar fields, or to coupling constants of field theory.
We may assign dimensions to the coefficients $g_\alpha$ as follows:
If we consider a $Dd$--brane  (or a d-dimensional part of
a higher dimensional brane) wrapped around a $d$-dimensional
supersymmetric cycle $C$ in the local geometry, it gives
a particle with BPS mass 
$$M=\int_C \Omega$$
which implies that the mass dimension of $[\Omega]=1$.
By quasihomogeneity, we can assign a dimension to
each variable $x_i$ proportional to its charge.  The
condition that $[\Omega]=1$ implies that the dimension
of a monomial $x_\alpha$ is given by
$$[x^\alpha]={2\over d-1-{\hat c}}\cdot Q_\alpha$$
and that the dimension of $g_\alpha$ is given by
\eqn\dimf{[g_\alpha]= {2(1-Q_\alpha)\over (d-1-\hat c)}.}
The deformation parameter 
$g_a$ is the expectation value of a canonically normalized
scalar field weighted with
the Yang--Mills coupling; that is, $g_a\equiv g_{YM}\langle\phi_a\rangle$. Since 
$[g_{YM}]= (4-D)/2 $ in $D$ spacetime dimensions, the unitarity bound
that $[\phi_\alpha]\ge (D-2)/2$ in a superconformal theory requires that 
%
$$[g_\alpha]={2(1-Q_\alpha)\over (d-1-\hat c)}>1$$
and this condition is identical to \normc .
Thus, the mode is a physical field
if and only if it satisfies the unitarity bound. This
is quite a satisfactory result (which has also been obtained
independently in \ref\kuta{A.~Giveon and D.~Kutasov,
``Little string theory in a double scaling limit,''
hep-th/9909110; A.~Giveon, D.~Kutasov and O.~Pelc,
``Holography for non-critical superstrings,''
hep-th/9907178.}).

Let us now consider Calabi--Yau manifolds of various dimensions and
see what this condition translates into in each case.

\subsec{d=2}
In the case of complex dimension 2, the finite distance
condition
\fd\ $\hat c <1$ implies that $W$ is given by the LG
superpotential of $N=2$ minimal models, which gives the
usual ADE classification of $K3$ singularities.  Moreover
the condition of physical fields \normc\
is automatically satisfied for all elements in the ring
because all the elements have
$$Q_\alpha \leq \hat c<{\hat c+1\over 2}.$$
This is in accordance (in the type IIA case) with the fact that all these
deformations can be viewed as giving vev to scalars in 
commuting directions 
of the adjoint representation of the corresponding group.

\subsec{d=3}
For Calabi--Yau 3-fold singularities, the condition
of being at finite distance implies that $\hat c<2$.
If we consider type IIB strings in the presence of such
singularities we obtain an $N=2$ superconformal theory
in 4 dimensions. Examples of such $W$'s are provided by 
\eqn\tvar{W=F(x,y)+z^2+w^2}
where $F(x,y)$ is a quasihomogeneous function of $x,y$.
In this case the function $F(x,y)$ can be identified
\ref\klmvw{A. Klemm, W. Lerche, P. Mayr,
C. Vafa, and N. Warner,
``Self-dual Strings and N=2 Supersymmetric Field Theory,'' 
Nucl. Phys. {\bf B477} (1996) 746,  hep-th/9604034.}\ 
with a fivebrane with worldvolume $R^4\times 
\Sigma$, where $\Sigma$ is the Seiberg-Witten
Riemann surface $F(x,y)=0$.  In this case
the singular Calabi--Yau gets mapped to a
singular Riemann surface.  
A special case of this 
$$F(x,y)=x^n+y^2$$
corresponds to the Argyres--Douglas points that we
will study in more detail later in this paper.

There
are other singularitities which are also at finite
distance which are not of the form \tvar.  For 
example consider
$$W=x^3+y^3+z^3+w^{3N}$$
Type IIB string in the presence of such
a singularity describes a superconformal theory, which
does not have a description in terms of a Seiberg-Witten
Riemann surface.  Such singularities do arise
in certain gauge theories.  For example the above
singularity appears in a gauge theory with gauge
group $SU(3N)\times SU(2N)^2\times SU(N)^3$
with certain bi-fundamental matter dictated by the
affine $E_6$ Dynkin diagram 
\ref\kmv{S. Katz, P. Mayr, and C. Vafa, 
``Mirror Symmetry and Exact Solution of 4D N=2 Gauge Theories,''
Adv.\ Theor.\ Math.\ Phys. {\bf 1} (1998) 53-114, hep-th/9605154.}.

In the case of the threefold, the condition that
deformations $x^\alpha$ correspond
to normalizable forms is \normc 
$$Q_\alpha< \hat c/2.$$
Given the pairing of the ring elements 
and the fact that the sum of the charges of the pairs gives
$\hat c$, this means that the deformations that
correspond to expectation values of dynamical fields
are in one to one correspondence with the deformations
that correspond to parameters in the theory.  A natural
interpretation of this pairing is to note
that for each chiral operator in the $N=2$ superconformal
theory $\Phi_\alpha$ (whose lowest component is identified
with $\phi_\alpha$  where $\langle \phi_\alpha \rangle =
g_\alpha$), we can consider the $N=2$ superspace integral
$$S\rightarrow S+\int d^4x d^4\theta \ t_\alpha \Phi_\alpha$$
with $t_\alpha$ being identified with the dual deformations
of $W$. The form of these deformations implies that the mass
dimensions obey $[t_\alpha]+[g_\alpha]=2$.
In fact this is consistent with what we have
found for the dimension of the parameters.  Namely, since
the charges of dual deformations add up to $\hat c$
we deduce that the charge corresponding to the $t_\alpha$
deformation is $\hat c-Q_\alpha$ and using the 
dimension of the fields given by \dimf\ we see that
indeed $[t_\alpha]+[g_\alpha]=2$:
$$[g_\alpha]={2(1-Q_\alpha)\over (2-\hat c)}$$
$$[t_\alpha]={2(1-{\hat c}+Q_\alpha)\over (2-\hat c)}$$
$$[g_\alpha]+[t_\alpha]=2.$$

\subsec{d=4}
The case of local singularities of Calabi--Yau 4-folds
was studied in \gvw. In particular, the case of theories
with $\hat c<1$ was studied in detail and it was shown
that type IIA string in the presence
of these singularities gives rise to certain
$N=2$ superconformal Kazama-Suzuki models in 2 dimensions.
In these cases none of the deformations correspond to 
dynamical fields
in the 2d theory.

\newsec{Perturbative String Description}
It is natural to consider the
perturbative string theory in
the presence of such singularities.  Of course, if we
are at the singularity the string theory is singular and
there is no perturbative expansion.  However, if we deform
the quasihomogeneous function $W$ we can resolve the singularity.
In such a case (if the singularity is resolved  enough so that
the wrapped branes are heavy enough) we can expect to have a
perturbative string description.  In this case one can use
ideas developed in \ref\gvwa{B. Greene, C. Vafa and N. Warner,
``Calabi--Yau Manifolds and Renormalization Group Flows,''
Nucl. Phys. {\bf B324} (1989) 371.}\ref\martg{E. Martinec,
``Criticality, Catastrophes, and Compactifications,'' 
in {\it Physics and Mathematics of Strings}, ed. L. Brink,
D. Friedan and A.M. Polyakov (World Scientific, 1990).}\
in the context of compact Calabi--Yau manifolds and 
for the non-compact case in 
\ref\muv{S. Mukhi and  C. Vafa,
``Two-Dimensional Black Hole as a Topological Coset Model of $c$=1 String 
Theory,'' 
Nucl.\ Phys.\ {\bf B407} (1993) 667-705, hep-th/9301083.}\ref\gov{D. 
Ghoshal and C. Vafa, ``$c$=1 String as the Topological Theory of the
Conifold,'' 
Nucl.\ Phys.\ {\bf B453} (1995) 121-128, hep-th/9506122.}\ref\oov{H. Ooguri and C. Vafa, 
``Two-Dimensional Black Hole and Singularities of CY Manifolds,''
Nucl.\ Phys.\ {\bf B463} (1996) 55-72, hep-th/9511164.}\ to find
a Landau--Ginzburg description of the theory. Namely
consider $d+2$ chiral fields $x_1,..,x_{d+1},y$
with $N{=}2$ $U(1)$ charge given by $q_1,...,q_{d+1},-1/\gamma$, and 
with a quasihomogeneous LG superpotential
\eqn\quasi{{\hat W}=W(x_i)+\sum_{\alpha} g_{\alpha}x^{\alpha}y^{\gamma (Q_\alpha
-1)}.}
$\gamma$ is fixed by the requirement that the total
${\hat c}=d$, which implies
that
$$\gamma={2\over d-\hat c-1}.$$
Note that the condition that the singularity be
at finite distance \fd\ is that $\gamma >0$ and the condition
that the deformation given by $g_\alpha$ corresponds to
a normalizable field in uncompactified theory is translated
to the condition that the power of $y$ accompanying that
deformation be less than $-1$.  To obtain the full supersymmetric
theory we also need to do the Gepner projection, which is a
discrete orbifold of the above LG theory that keeps
only the integral $U(1)$ charged fields.

A particular case of \quasi\  is 
the deformation of $W$ by a constant:
$$\hat W=W+g_0 y^{-\gamma}$$
In this case it can be shown \muv \gov \oov\ 
that the resulting theory corresponds to the tensor product
of the LG theory given by $W$ and a Kazama-Suzuki
coset model $SL(2)/U(1)$ with level $k=2+\gamma$.
This construction has also been noted independently in
\kuta .  Certain aspects of conformal theories near
singularities have also been discussed recently in
\ref\song{J.S.~Song, ``Three-dimensional Gorenstein 
singularities and SU(3) modular  invariants,''
hep-th/9908008.}.

\newsec{Solitons in the $d=3$ case}
For the rest of this paper we will  be interested
in studying some aspects of Calabi--Yau threefold
singularities corresponding to Argyres--Douglas points.
As already noted these points correspond to local 
singularities given by
\eqn\wpx{W=P(x)+y^2+z^2+w^2=0}
with $P(x)$ a polynomial of degree $n$ in $x$.
The superconformal point itself corresponds to $P(x)=x^n$, 
and there are $n-1 $ deformations given by
$$P(x)=x^n+\sum_{i=0}^{n-2}g_i x^i$$
where  the $g_i$ correspond to expectation value of fields
for $i<(n-2)/2$ and to dual mass parameters
for $i>(n-2)/2$ \ref\apsw{P.C. Argyres, M.R. Plesser, N. Seiberg, and
E. Witten, ``New N=2 Superconformal Field Theories in Four Dimensions,''
Nucl.\ Phys.\ {\bf B461}, 71 (1996)
hep-th/9511154.}. 
Eq.\centc\ gives the central charge
$$ \hat c = 1- {2\over n} $$
and \dimf\ gives the dimensions 
$$ [g_i]={2(1-i/n)\over 1+ 2/n}={2(n-i)\over n+2}\qquad\qquad (i=0..n{-}2)$$
in agreement with the result of 
\ref\ehiy{T. Eguchi, K. Hori, K. Ito, and S.-K. Yang,
``Study of N=2 Superconformal Field Theories in Four  Dimensions, Nucl.\
Phys.\ {\bf B471} (1996) 430, hep-th/9603002.}. 

We are interested in studying type IIB 
strings in the presence of such singularities.
In particular, we would like to study the BPS states in
this theory near the superconformal point and how their
spectrum jumps
as we change the parameters $g_i$ in the polynomial $P(x)$.

The BPS states in this case correspond to D3--branes 
wrapped around supersymmetric
3-cycles. A basis of vanishing 3-cycles (not
necessarily supersymmetric) can be chosen to be
$n{-}1$ 3-spheres intersecting one another
according to the Dynkin diagram of $A_{n-1}$.
The intersection of two cycles can be interpreted as the
skew symmetric form in the product of electric and magnetic
charges.  In particular, two $S^3$'s corresponding
to adjacent nodes of $A_{n-1}$ carry, in some basis,
electric versus magnetic charge.

To find the supersymmetric 3-cycles, one follows the
strategy in \klmvw\ and considers the 3-cycle
as a 3-sphere  consisting of an  $S^2$ fibered
over a real curve in the $x$-plane.  The $S^2$ above a given
$x$ is $y^2+z^2+w^2=-P(x)$ (in an appropriate real subspace). 
The projection of the 3-cycle onto the $x$-plane is 
a curve in the $x$-plane which begins and ends
at zeroes of $P(x)$. At the endpoints of the curve, 
which correspond to the `poles' of $S^3$, the
radius of the $S^2$ in the fiber goes to zero. 
A basis of 3-cycles
(not necessarily supersymmetric) can be chosen
to correspond to $n-1$ intervals connecting the $n$ zeroes
of $P(x)$ in a sequence.  These would correspond to a set of 
3-cycles whose intersection gives the Dynkin diagram of $A_{n-1}$.

As discussed in \gvwa\ the condition of having
a supersymmetric cycle  gets translated
in the $x$-plane into the existence 
paths beginning and ending
at the zeroes of $P(x)$ such that the phase of
\eqn\pdx{\int_{S^2} \Omega =
\sqrt{P(x)}\,dx}
is constant along the path. This would guarantee in particular
that the BPS inequality 
$$\int_{S^3} |\Omega |\ge |\int_{S^3} \Omega |$$
is saturated.
Alternatively, the required condition
is that the image of the path under the 
Jacobian map with respect to the reduced one--form
should be a straight line in the flat $W$-plane, that is, 
\eqn\jac{
W(x(t))=\int_{x_0}^{x(t)} \sqrt{P(x)}\,dx=\alpha t }
where $t$ is real parameterizing the path
and $\alpha$ is a phase. In this case, the mass of a D3--brane 
wrapped around the supersymmetric cycle 
corresponding to such a path between zeros at $x_0$
and $x_1$ automatically saturates the BPS inequality
$$
M=\int_{x_0}^{x_1}\left| \sqrt{P}\, dx\right| 
\ge \left| \int_{x_0}^{x_1}\sqrt{P}\, dx\right| 
$$

An efficient technique to find such paths is to 
solve the first--order differential equation
$$
\sqrt{P(x)}\,{dx\over dt} = \alpha.
$$
for solutions $x(t)$ beginning and ending at roots of $P(x)$,
for all possible values of the phase of $\alpha$. 
This is equivalent to finding integral curves of the vector
field 
\eqn\vf{
{\alpha\over \sqrt{P(x)}}\,{\partial\over \partial x}
}
that start and end at roots of $P$. As we have explained, the existence 
or nonexistence
of such a curve then implies the existence or nonexistence of
a particular BPS state. This procedure is easily implemented. 

In the next section we analyze
the solutions to the condition \jac\ in various regimes of
parameters.

\newsec{Finding the BPS states}
As we have discussed, we need to find solutions to
\eqn\maineq{{dx\over dt}={\alpha \over \sqrt{P(x)}}}
where $t$ is a real parameter and $\alpha$ is a phase.  Moreover
the solution should begin and end at one of the $n$ roots of $P(x)$.
We need to establish a few facts:

{\it For each topology of path between two 
roots of $P(x)$}, $\alpha$ {\it is uniquely fixed} (up to an overall sign).
By topology of path we mean a homology class 
of paths with fixed endpoints
on the complex $x$-plane with all the roots of $P(x)$ deleted.  
To see the uniqueness of $\alpha$ note that
for each path $\gamma$ in a particular homology class
the integral
\eqn\period{\int_{\gamma} dx \sqrt{P(x)}=a_{\gamma}.}
which is well defined up to an overall sign,
is independent of the precise choice of $\gamma$
in that class.  This is because two different paths
in the same class will differ by integral of an analytic
expression in the region bounded by the two curves.  Physically,
this is simply the reflection of the fact that each choice of
path topology fixes the electric and magnetic charges of
the BPS state and the integral \period\ is just the central
charge of the $N=2$ algebra in that sector.
It follows that $\alpha$ is uniquely fixed for each
choice of class of path to be 
$$\alpha=a_{\gamma}/|a_{\gamma}|.$$

The next fact we need to establish is that 
{\it there is at most one solution connecting two 
roots}.
We first show that there
cannot be two different solutions in the same path class.
Let us first analyze the structure of the solution to \maineq\
near each root.  Near a root of $P(x)$, which with no
loss of generality we take to be at $x=0$,
one is solving
an equation which can be approximated as
$${dx\over dt}={\alpha\over \sqrt{x}}$$
whose solutions are given by
$$x=(\hbox{$3\over 2$}\alpha t)^{2\over 3}$$
In particular, for a given $\alpha$ there are {\it three}
solutions near $x=0$ which make 120$^\circ$ angles relative
to one another, corresponding to the 3 possible choices of 
cube roots of $\alpha^2$ in this solution. A typical
set of integral curves near to a root is depicted in Figure 1.

\centerline{\epsfxsize 3truein \epsfysize 2.8truein\epsfbox{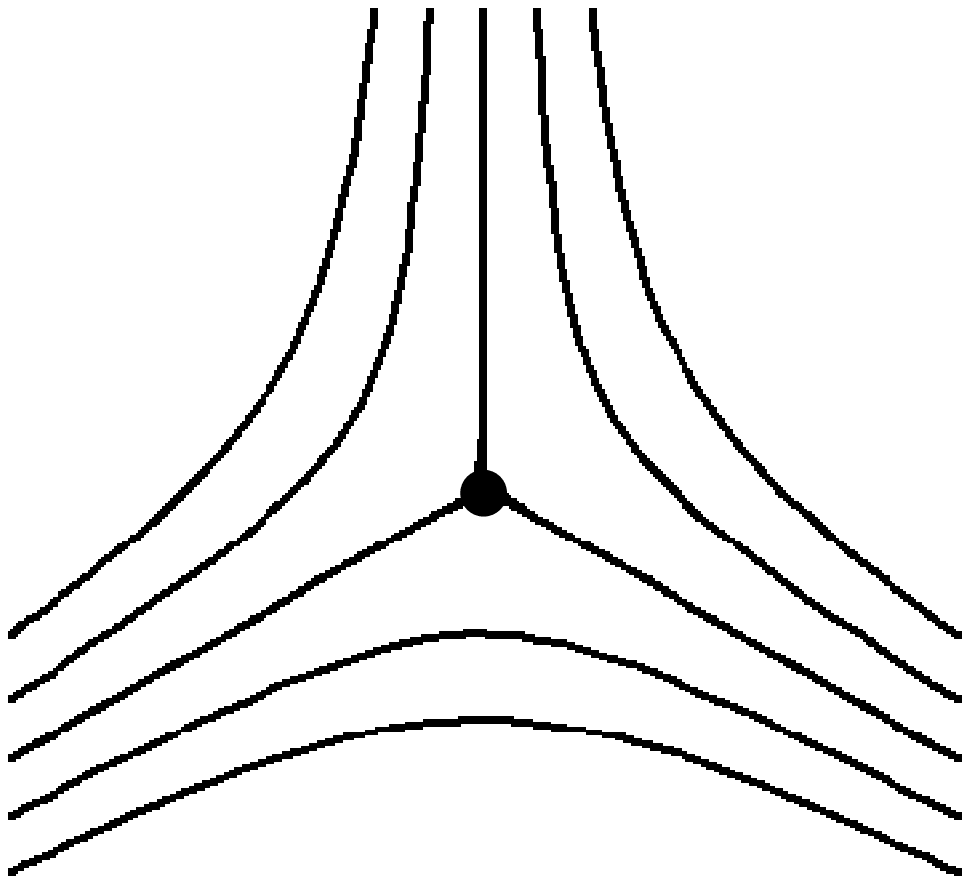}}
\noindent{\ninepoint\sl \baselineskip=8pt {\bf Fig.1}: {\rm
A typical set of integral curves near to a root.}}
\bigskip

Now suppose there are two
solutions in a given class (and thus with the same choice of $\alpha$)
connecting two roots of $P$.  In other words, let us assume
that two of the three integral curves from one root
join with two of the three integral curves from
another root (see Figure 2a).  The geometry
should be as depicted in Figure 2a:  first, the paths cannot intersect
each other except at the two ends, 
as the vector field has a well defined direction
at each point and that would not be the case at an intersection point.
Secondly, the paths must make an angle of $120^\circ$ on each
side, otherwise the third integral curve emanating from either of the roots
would have nowhere to go. (We can also rule out the case 
where all three curves are joined,
by applying this argument to two of the curves making 
$120^\circ$ angles relative to each other.) From 
the geometry of  Figure 2a it is clear that
there will necessarily
be closed integral curves trapped inside.  If we consider
the winding of the phase
$$\Phi =|\sqrt{P(x)}|/\sqrt{P(x)}$$
defined by
$$w={1\over 2 \pi i}\int d\, \log\Phi$$
we see that $w=+1$ along these curves.
Since $P(x)$ has no zeroes or poles inside this curve, this is impossible.
We could also have chosen, instead of the trapped integral
curve, an arbitrary interior curve to use for counting the
winding of the phase $\Phi$.  For example
we can make the choice shown in Figure 2b.  In this
case we again get winding number +1, with the two small arcs
near the roots each contributing $+1/6$ to the winding
number and the remaining pieces along the integral curves giving
a net winding of $+2/3$.  

\centerline{\epsfxsize 4truein \epsfysize 3.3truein\epsfbox{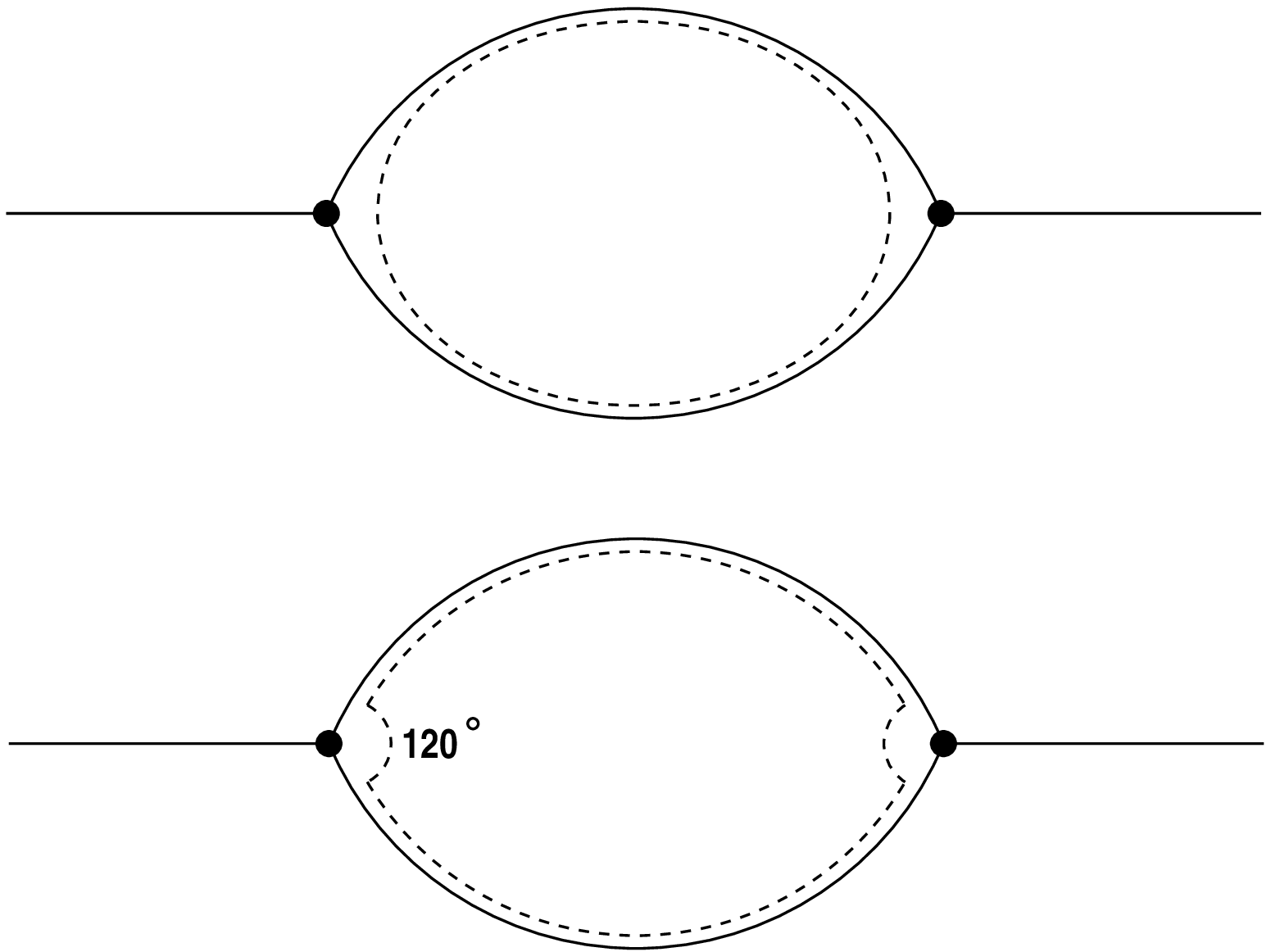}}
\noindent{\ninepoint\sl \baselineskip=8pt {\bf Figs.2a and 2b}: {\rm
Impossibility of more than one integral curve of a given topology
between two roots.
The winding of the phase is calculated by integrating along the dashed line.}}
\bigskip

Now let us also establish that there can not be two solutions
between two roots, even if the paths are in different
topological classes. In such a case the two paths
correspond to solutions to \maineq\ with two different phases
$\alpha_1$ and $\alpha_2$.  Let $\epsilon= \alpha_1\alpha_2^{-1}$
and define $\delta\equiv{1\over 2\pi i} \log \epsilon$.
With no loss of generality we can assume $\epsilon$
to be some generic phase (in other words for any special values of
$\epsilon$ we can change the coefficients of $P(x)$
so that the two central
charges have infinitesimally different phase ratios).
Let us consider the winding
of the phase $\Phi(x)$ along the closed path
shown in Figure 3. We will first assume, as in Figure 3, that
the two curves do not intersect except at the endpoints.
This assumption will be justified shortly.

\centerline{\epsfxsize 2.5truein \epsfysize 2.2truein\epsfbox{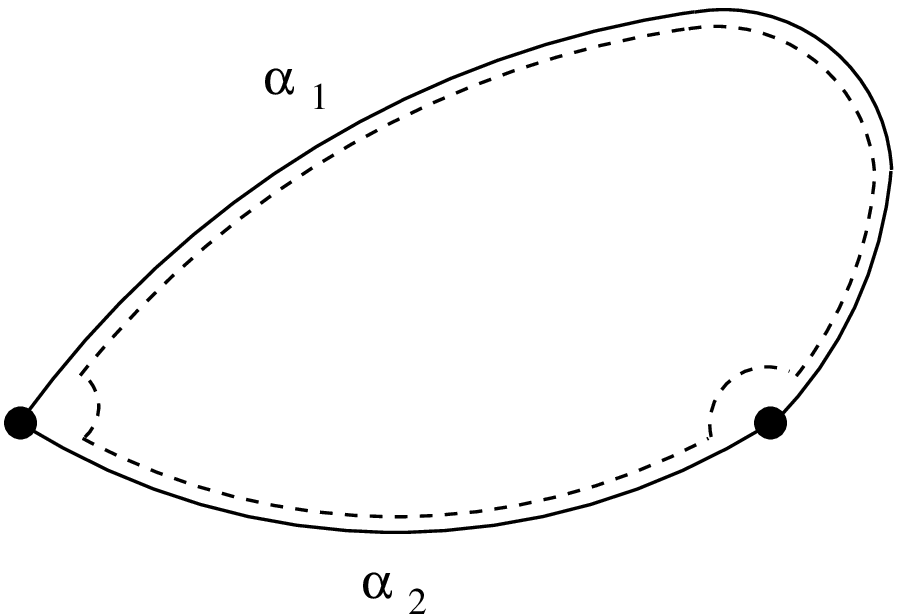}}
\noindent{\ninepoint\sl \baselineskip=8pt {\bf Fig.3}: {\rm
Two integral curves between roots, with $\alpha_1\ne \alpha_2$.}}
\bigskip

\noindent 
This closed path consists of the two integral curves and the small
paths connecting them near the roots.  Even though the two
integral curves are for different choices of $\alpha$, in
either case $\alpha$ is a constant and does not vary over the
path. In particular the integral $\int d\log\Phi$ along the two integral 
curves is the same as winding of the velocity vector, 
and gives a total contribution to the
winding of $+2/3$. Along the arcs near each of the roots, the
contribution to the winding will be given by $1/6$ up to a
correction by  $\delta/6$ at one endpoint and
$-\delta/6$ at the other endpoint, which cancel out.  
We thus still obtain a net winding of $+1$ unit.  However this is impossible as the
winding associated with $1/\sqrt{P}$ is in general negative and given
by $-m/2$ where $m$ denotes the number of roots of $P$ inside the closed
curve.  The reason for this is that $1/\sqrt{x-a}$ has a winding number $-1/2$
around $x=a$. (The other possible geometries for these two integral curves
curves --- in which their relative angle of approach to one of the endpoints
is shifted by $\pm 2\pi/3$ --- lead to similar contradictions.)

\centerline{\epsfxsize 3truein \epsfysize 2truein\epsfbox{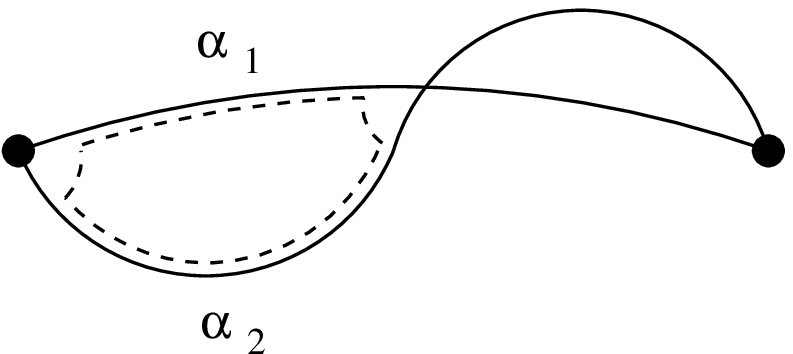}}
\noindent{\ninepoint\sl \baselineskip=8pt {\bf Fig.4}: {\rm
Impossibility of intersection of two integral curves.}}
\bigskip

We also need to justify our assumption that the curves
do not intersect each other except at the endpoints.  Suppose
they did.  Pick an endpoint and 
consider the closed curve that forms between it and 
the first point where the two integral
curves intersect (Figure 4).  The winding number around that closed curve
is again the sum of contributions of the two integral curves
and the small arcs near the sharp ends. The integral curves contribute 
the sum of the two angles divided by $2\pi$, or ${1\over 3}(1+\delta)+\delta$, 
the left endpoint adds ${1\over 6}(1+\delta)$, and the right endpoint adds nothing.
So the net winding number ${1\over 2}+{3\over 2}\delta$ is not even an integer, 
as $\epsilon$ can be assumed to be a generic phase.  This is clearly a contradiction
with the fact that the winding should be $-m/2$ for some integer $m$.

We have thus established that there is at most one solution
connecting two distinct roots. 

{\it Minimum Number of BPS states}: We can also argue
that any pair of roots are connected by
a sequence of BPS solutions.  We can find such a sequence by 
minimizing $\int |\sqrt{P(x)}dx|$ over arbitrary paths
beginning at one root and ending on the other.   
Clearly there will be a solution to the minimization
condition (paths going to infinity will give an infinite
contribution to the above integral).  Consider the path
which minimizes it. Then for a generic point on this path,
away from the roots of $P$, the curve should satisfy \maineq .
If not, fix two nearby points on the curve and take
a solution of \maineq\ between these two points.  This will
have a {\it lower} value for the integral, violating 
the assumption that we had found the curve which minimizes the
integral.  In general, this minimal curve may pass through some roots of $P$.
The above argument still shows that piecewise, between the roots,
it should satisfy \maineq\ for some $\alpha$.  This implies that any two
roots are connected by a sequence of BPS solutions.
If there are $n$ roots in all, the minimum number of BPS solutions is thus 
$n{-}1$. 

\centerline{\epsfxsize 6.5truein \epsfysize 3truein\epsfbox{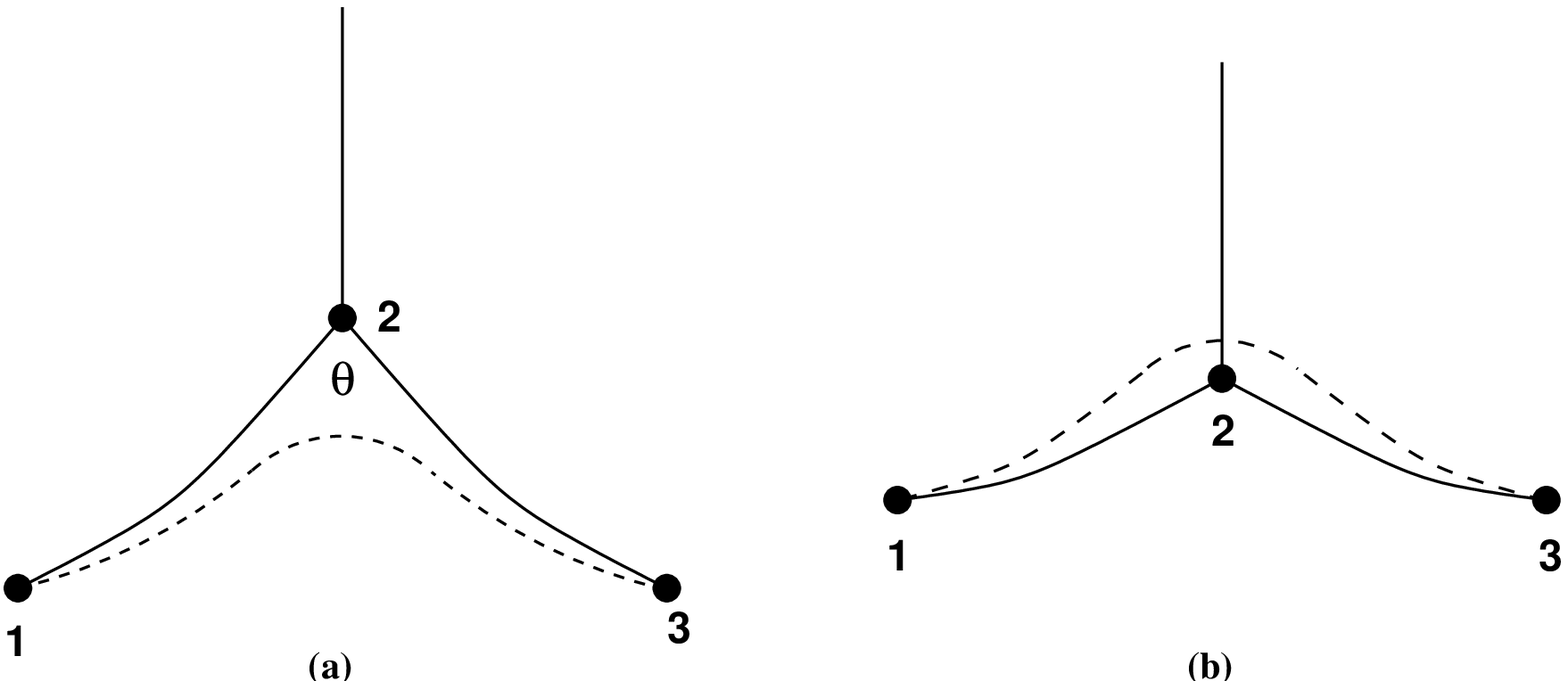}}
\noindent{{\ninepoint\sl \baselineskip=8pt {\bf Fig.5a}: \rm
Integral curve from 1 to 3 when $\theta<120^\circ$. {\bf Fig.5b}:
When $\theta>120^\circ$ no 1--3 curve exists.}}
\bigskip

{\it Jumping Phenomena}: Consider three roots of $P$ labeled by 1, 2,
and 3.  Suppose for some choice of parameters there are BPS solutions
from 1 to 2, from 2 to 3, and from 1 to 3.  {\it A priori}, the phase of
the central charge of the $N=2$ algebra, which determines the phase
$\alpha$, will be different for the three different solitons.  However,
let us assume that as we change the parameters the phases become the
same, \ie, the BPS charges from 1 to 2 and 2 to 3 get aligned and
sum up to the charge from 1 to 3.  This means that for the {\it same}
choice of $\alpha$, i.e. $\alpha_{12}={\alpha}_{23}={\alpha}_{13}$, we
have three integral solutions.  In such a case the curve from 1 to 3
must coincide with the concatenation of the curves 
from 1 to 2 and 2 to 3. 
If not, as in the
previous arguments, by considering the trapped integral curves inside we
would get a contradiction with the winding number.  At the alignment
point the angle
$$\theta =\widehat{123}$$
which gives the angle between the solitons 1--2 and 2--3 is 120
degrees. However as we change parameters from one side to the other
side the angle changes from less than to greater than 120$^\circ$. We will
argue below that the case where the 1--3 soliton does exist corresponds to
when $\theta <120^\circ$). First, we  argue that the 1--3 soliton should
disappear when $\theta >120^\circ$.  If the 1--3 solution continued to
exist, the case before (when $\theta< 120^\circ$) and after
($\theta>120^\circ$) the transition should look like that depicted in
Figures 5a and 5b.  This is because the phase of 
$\alpha_{12}/\alpha_{13}$ should go from one sign to another, which
implies, by looking at integral curves near the point 1, that the 1--3
curve is to one side or the other of the 1--2 curve.  But the
curve 1--3 depicted in Figure 5b is forbidden: The third integral
curve emanating from point 2 would intersect it, which is not allowed.  We thus
see that the soliton represented by the 1--3 curve decays to 1--2 and
2--3 soliton as we pass through alignment.

\centerline{\epsfxsize 3truein \epsfysize 3truein\epsfbox{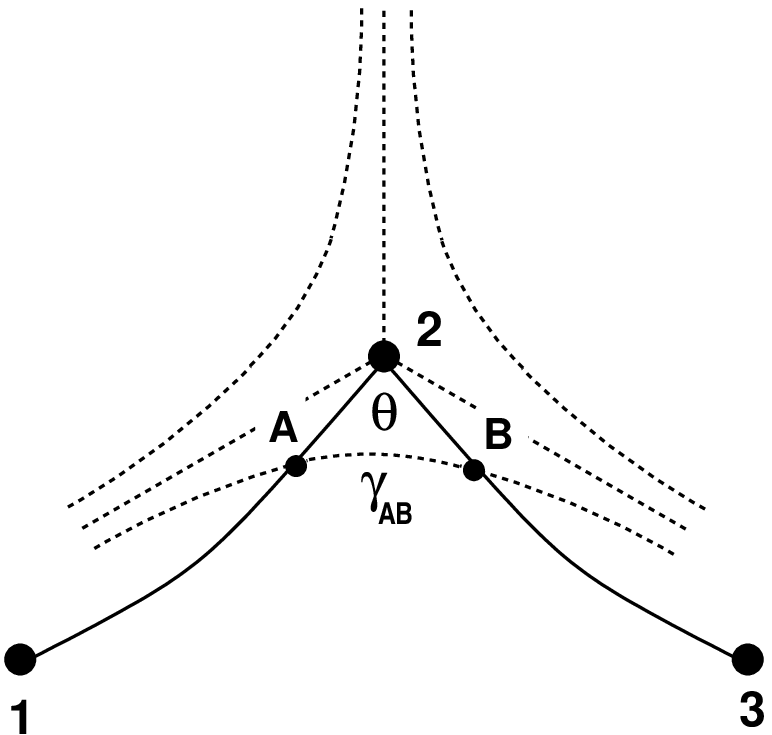}}
\noindent{\ninepoint\sl \baselineskip=8pt {\bf Fig.6}: {\rm
Proof of existence of integral curve from 1 to 3  when $\theta<120^\circ$.}}
\bigskip

The reverse of this
argument can also be made.  Suppose we originally have no soliton from
1 to 3.  Let us choose parameters such that we are close to an aligned
configuration, which as argued should correspond to $\theta =120^\circ$.
Then we argue that as we decrease $\theta$ a soliton should
appear from 1 to 3.  To see this note, as we discussed already, that there
is a solution which minimizes the integral $\int |\sqrt{P}dx|$ between
any pairs of point.  Apply this to points 1 and 3.  At the alignment
point the minimal solution is given by the sum of the curves 1--2 and 2--3.
If $\theta$ is decreased to less than $120^\circ$ the class
minimizing it is no longer sum of the 1--2 and 2--3 curves. The integral
can be decreased, for example by considering a path such as the one
shown in Figure 6: we pick two points $A$ and $B$, on 1--2 and 2--3
curves respectively, near root 2.  We consider a
solution of the integral curves $\gamma_{AB}$ of the vector field
which interpolates between $A$ and $B$ for some $\alpha$.  That this is
possible is guaranteed if $\theta <120^\circ$.  Then the integral $\int
|\sqrt{P}\,dx|$ is decreased if for the part near point 2,
the curve $\gamma_{AB}$ is used, instead of the radial lines
connecting $A$ and $B$ to point 2.  Thus the minimum does not coincide
with the sum of the 1--2 and 2--3 curves, 
and there  appears a distinct 1--3 soliton as in Figure 5.  The
local analysis here is similar to what Joyce has considered in the
context of 3-cycles on Calabi--Yau \ref\joyce{D.~Joyce,
``On counting special Lagrangian homology 3-spheres,''
hep-th/9907013.}\
\foot{We would like to thank M. Douglas for pointing
this reference out to us. See also \ref\kachru{S. Kachru
and John McGreevy, ``Supersymmetric Three Cycles
and Supersymmetry Breaking,'' hep-th/9908135.}
\ref\douglas{M.R. Douglas, ``Topics in $D$ Geometry,'' hep-th/9910170.}.}. 

To summarize, whenever the phases of 3 such BPS charges are 
equal, we are in a situation of marginal stability. Under a perturbation 
away from alignment in one direction, all 3 states will be stable, 
but in the other direction, one of the states will become unstable to 
decay into the other two states.

\subsec{Examples}

In this part we show that with suitable choices of $P(x)$ we can have
any number of solutions between $n-1$ and $n(n-1)/2$.  That $n-1$ is
the minimum possible number has already been noted.  Also that
$n(n-1)/2$ is the maximum number follows from the fact that there is
at most one BPS state for each pair of roots of $P$.  We will show
that, for example, with $P(x)=x^n-1$ we get $n(n-1)/2$ solitons, one for
each pair of roots of $P$.  On the other hand with $P(x)$ having only
real roots, we show that $P$ has exactly $n-1$ solitons, corresponding
to the solitons connecting adjacent roots.  By our discussion about
the jumping phenomenon, it follows that as we continuously change the
polynomial $P(x)$ we get arbitrary number of solitons anywhere between
these two bounds.  In this way the story is very similar to that of
the $A_n$ series for $N=2$ LG theories in 2 dimensions \foot{In fact
if we consider a degenerate choice of polynomial, where $P=(dW/dx)^2$
the problem of solving \maineq\ is identical to the problem of finding
solutions to 2d solitons in an $N=2$ theory with LG potential given by
$W$.}.

Let us first focus on the simplest nontrivial $A_n$ singularity, with
$P(x)=x^3$, corresponding to the original SU(3) Argyres--Douglas point \ad. 
By the above arguments, the minimum possible number of BPS states is 
$n-1=2$, and the maximum is 3.
A general deformation of $P(x)$
$$P(x)=x^3+ux+v$$
involves two parameters $u$ and $v$, with $v$ an operator expectation
value and $u$ a mass parameter.  As we shall describe, the stable BPS
spectrum near the singularity depends on the direction of approach,
and the marginal stability surfaces (MSS's) extend all the way down to
the critical point.  By tuning the dimensionless ratio $u^3/v^2$
appropriately as the scaling is performed, we may end up with either
two or three light stable BPS states. (This modifies the result of
\gh, which found three states.) We will now exhibit examples in which 
each of these possibilities is realized.

First consider a situation where all three roots are arranged
symmetrically, at cube roots of unity times a common scale factor, so
that $P(x)=x^3-1$.  Given any integral curve linking two roots at some
value of $\alpha$, we can construct two others, linking the other two
pairs of roots, by multiplying $\alpha$ and $x$ by an appropriate cube
root of unity (which preserves $P(x)$ and the form of Eq.\jac). Hence,
the total number of integral curves must be at least 3. Since the total
number of such curves must also be either 2 or 3, this maximally
symmetric configuration of roots leads to exactly 3 stable BPS states.

As another example, 
suppose the roots are all colinear; without loss of generality,
we may suppose that
\eqn\plambda{
P(x)=x(x-1)(x-\lambda)}
where $\lambda$ is real. We claim that there are exactly two integral curves
for all $\lambda >1$. By symmetry, all integral curves must either be at
$\alpha=\pm 1$ or $\alpha=\pm i$, and must lie along the real axis.
Otherwise, given an integral curve not satisfying these conditions, we could
construct another inequivalent geodesic between the same two roots, at
the conjugate value of $\alpha$ and along a conjugate path, in
contradiction to the general uniqueness argument above.  (Note that
solutions at $\alpha=i$ and $\alpha=-i$ should be considered
equivalent; the change in sign of $\alpha$ simply means that the path
is traversed in the opposite direction.) Now it is clear that there
can be no integral curve from $x=0$ to $x=\lambda$, because it would have to
pass either over or under $x=1$, and thus could not lie entirely along
the real axis.  Thus, there must be exactly two integral curves (since this
is the minimum possible), from 0 to 1 (which occurs at $\alpha=1$) and
from 1 to $\lambda$ (at $\alpha=i$). These correspond to real and
imaginary 3-spheres in the Calabi--Yau manifold, intersecting
transversely at a point.

Next, consider interpolating between these configurations by
holding two of the roots fixed 
and moving the third root
along a line midway between them. We do this by varying $\lambda$ in 
Eq.\plambda\ from ${1\over 2} + i{\sqrt3\over 2}$ to ${1\over 2}$,
along the line Re$(x)={1\over 2}$. 
We find numerically that at approximately $\lambda={1\over 2}+.231 i$, the 
number of integral curves jumps from 3 to 2. The curve of marginal stability 
in the $\lambda$--plane has roughly the appearance shown in 
Figure 7, and is preserved by the modular transformations 
\eqn\modular{
\lambda \longrightarrow {1\over \lambda}, 
\qquad \lambda\longrightarrow 1-\lambda, 
\qquad \lambda\longrightarrow\bar\lambda.
}
In each of 
the regions containing the real $\lambda$--axis, 
the number of stable BPS states is 2; 
elsewhere in the $\lambda$-plane, there are 3 such states. 
By a local analysis, it can be shown that the components of the curve
intersect the real axis at angles of $\pm 60^\circ$, and approach
asymptotes also making angles of $\pm 60^\circ$.

\centerline{\epsfxsize 4truein \epsfysize 3truein\epsfbox{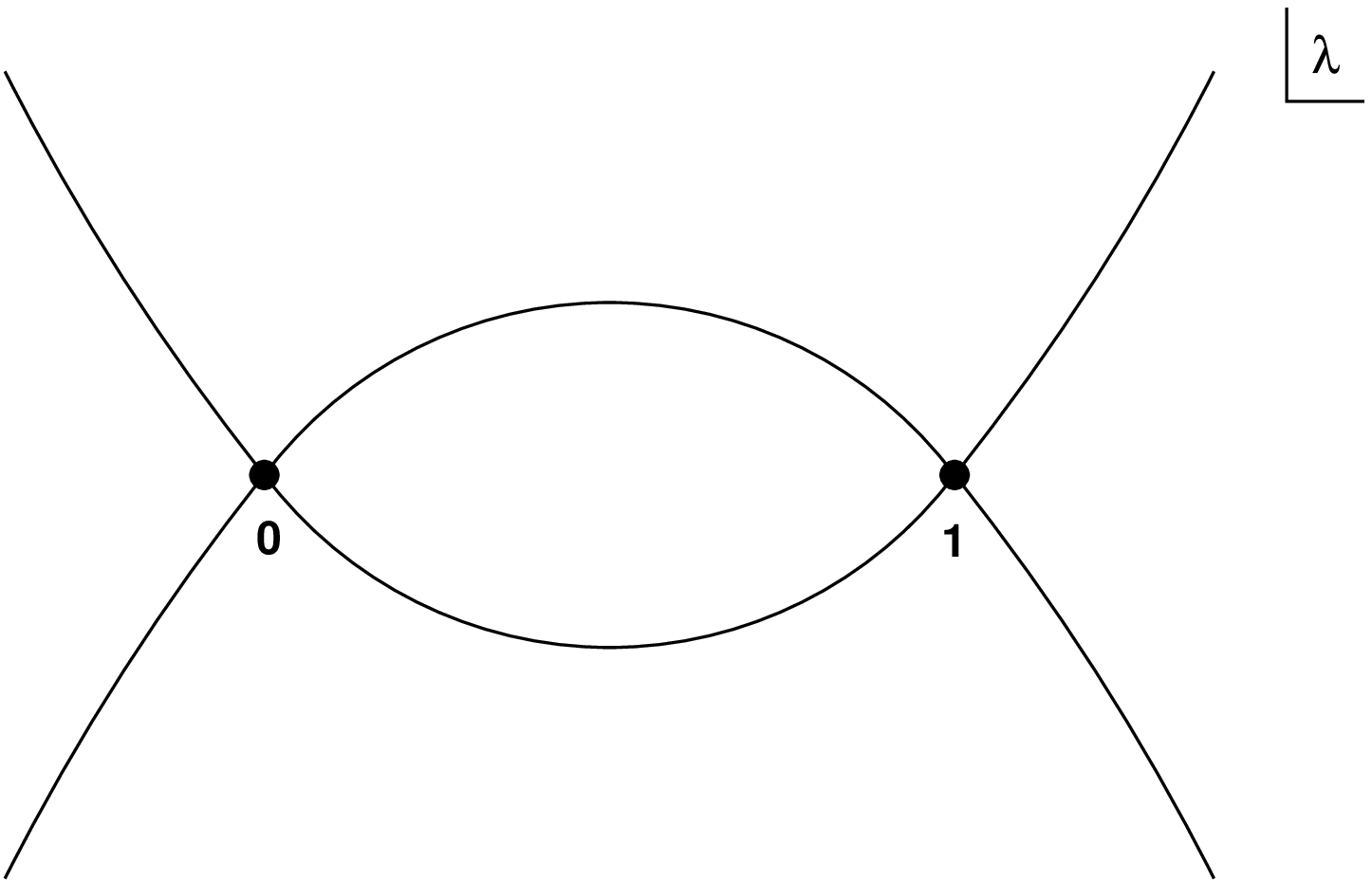}}
\noindent{\ninepoint\sl \baselineskip=8pt {\bf Fig.7}: {\rm
Curves of marginal stability for $n=3$.}}
\bigskip

The general $A_n$ singularity with $P(x)$ an $n$th order polynomial
may be analyzed by the same methods.  Again, when the roots are
arranged with $\Z_n$ symmetry, we will show that the maximum possible
number $n(n-1)/2$ of stable BPS states is realized.  And when the
roots are colinear, the number of stable BPS states is
minimized. Indeed, in the latter case, the same argument we gave for
$n=3$ extends to show that there are precisely $n-1$ integral curves,
connecting the roots in sequence, with $\alpha$ alternating between
$1$ and $i$. 

On the other hand, suppose that the roots are located at the $n$th roots
of unity, corresponding to the polynomial
$$ P(x)=x^n-1 ~,$$
We will now show that there is a unique integral curve
connecting any two roots, for a total of $n(n-1)/2$
stable hypermultiplet states. 

Assume for simplicity that $n$ is even.
In general, all integral curves of Eq.\maineq\ 
for an $n$th--order $P(x)$ that extend 
out to infinity in the $x$-plane
must asymptote to one of $n+2$ lines of constant phase. 
This is because, for large $x$, Eq.\maineq\ 
has the asymptotic solution
$$
x(t)=((n+2)\alpha t)^{1/(n+2)}. 
$$
As we have discussed, there are exactly 3 integral curves emanating
from each root; each of these must end either at another root or at
one of these $n+2$ asymptotic infinities. We consider the graph formed
by the set of all integral curves starting or ending at a root, such
as the graph shown in 
Figure 8 for the case $n=6$ and $\alpha=1$.

\bigskip
\centerline{\epsfxsize 4truein \epsfysize 4.5truein\epsfbox{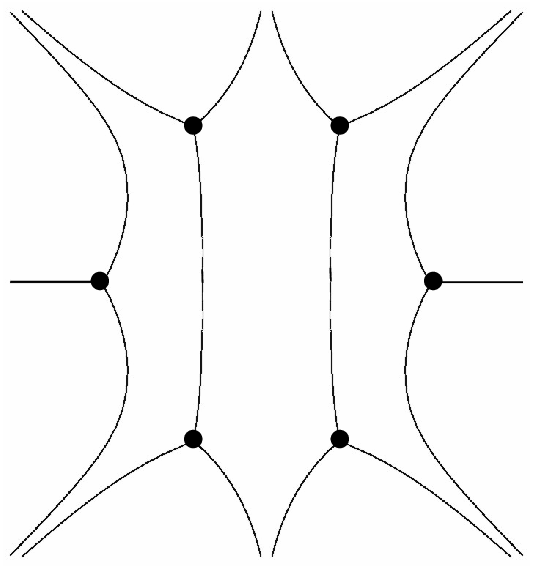}}
\noindent{\ninepoint\sl \baselineskip=8pt {\bf Fig.8}: {\rm
Graph of integral curves ending on roots for $n=6$ and $\alpha=1$. }}
\bigskip

An important fact, which severely constrains the topologies of allowed
graphs of this type, is that {\it no two integral curves from a single
root may approach the same asymptotic infinity}. The proof is almost
identical to the above argument that no two roots can be joined by two
distinct integral curves. If this did happen, we could consider the
concatenation of two integral curves between a given root and a given
infinity.  A small deformation would produce a curve beginning and
ending at the same infinity, with net index of $+1/2$. But as we have
seen, closed curves always lead to indices of $0$ or less, a
contradiction.

We now argue that, for $\alpha{=}1$, there are integral curves
connecting complex--conjugate $n$th roots of 1, and all other curves
in the graph are unbounded. To see that only conjugate points can be
connected, suppose that two non--conjugate roots
$x_j= \exp(2\pi ij/(n+2))$ and $x_{j+k}=\exp(2\pi i(j+k)/(n+2))$,
were joined by an integral curve.  
Then complex conjugation would map this curve into a different
integral curve, also with $\alpha{=}1$,
between a pair of roots conjugate to the original pair. 
By $\Z_n$ symmetry, at other
values $\alpha=\exp(2\pi ik/n)$, we obtain $n$ rotated pairs of
integral curves. Generically, this procedure leads to multiple
geodesics between $x_j$ and $x_{j+k}$, at distinct values of $\alpha$.
(A special case, when the original two roots are related by reflection
about the imaginary axis, can be ruled out by showing, as we do below, 
that integral curves joining such points already exist for $\alpha=i$.)

Next, we need to show that, for $\alpha=1$, all conjugate roots are
indeed connected by integral curves. We first assume that $n$ is even.
The roots $x=1$ and $x=-1$ are
special; they are not connected to any other roots, and each of them
has 3 integral curves which must terminate at 3 different asymptotic
infinities, as seen in Figure 8.  
By complex conjugation symmetry, one of the integral
curves from $x=1$ lies entirely along the real axis; the other two are
asymptotically parallel to the lines $x(t)=\exp(\pm 2\pi i/(n+2))t$.  In
traversing a very large circle clockwise in the $x$--plane, the
winding index of any vector field defined by \maineq\ will be
$(n+2)/2$. As we traverse the circle clockwise from the point where
the first integral curve from $x=1$ crosses this circle to the point where
the third curve crosses, the index shifts by at least $+1$.  Now
consider any root connected to its conjugate; it must also have two
unbounded integral curves, which similarly contribute at least $+1/2$
winding to the total index. If all conjugate roots are connected, the
total index will be at least $(n+2)/2$; if any are not connected, their
individual contributions to the index will be at least $+1$, and 
the lower bound on the index will be greater than its actual
value. Therefore, for $\alpha=1$, all conjugate roots are connected,
and we have thus shown the existence of ${n\over 2}-1$ integral curves
of finite length.  The graph of integral curves for the case $n=8$ 
is shown in Figure 8.

The same sort of argument holds for each $\alpha$ which is an $n$th root of 1.
In each of the $n/2$ cases (\ie, not distinguishing between $\alpha$
and $-\alpha$), we can exploit the symmetry about the line $\alpha t$
to obtain ${n\over 2}-1$ integral curves, for a total of ${n\over
2}({n\over 2}-1)$ distinct cases. We can also make use of the symmetry
about the line $\alpha t$ when $\alpha$ is a $(2n)$th root of 1; this
gives an additional $({n\over 2})^2$ integral curves, for a total of
$n(n-1)/2$.  Finally, if $n$ is odd, a similar counting scheme works: each
$n$th root of 1 corresponds to a distinct symmetry axis, and each such 
value of $\alpha$ leads to $(n-1)/2$ integral curves, for a grand total of 
$n(n-1)/2$. 

\bigskip
{\bf Acknowledgements:}
We are grateful to P. Argyres, M. Crescimanno, M. Douglas, and D. Kutasov 
for useful discussions. AS thanks Harvard University and CV thanks
Rutgers University, for hospitality while this work was in progress.
The research of AS is supported by NSF Grant No. PHY--9722147;
CV is supported by NSF Grant No. PHY--9218167. 

\vfill\eject

\listrefs

\end